\documentstyle[aps,prb,multicol]{revtex}
\input{psfig}
\begin{document}
\newcommand{\lee}{{l_{ee}}}
\newcommand{\vecr}{{\bf r}}
\newcommand{\vecy}{{\bf y}}
\newcommand{\veck}{{\bf k}}
\newcommand{\vecv}{{\bf v}}
\newcommand{\vecE}{{\bf E}}
\newcommand{\vecj}{{\bf j}}
\newcommand{\kperp}{{{\bf k}_\perp}}
\newcommand{\sinch}{{\rm sinch}}
\newcommand{\cth}{{\rm cth}}
\newcommand{\bartau}{{{\bar \tau}_T}}
\renewcommand{\Re}{{\rm Re}}
\newcommand{\spc}{{\,\,\,\,\,\,\,\,}}
\newcommand{\bea}{\begin{eqnarray}}
\newcommand{\eea}{\end{eqnarray}}
\renewcommand{\[}{\begin{equation}} 
\renewcommand{\]}{\end{equation}}
\newcommand{\bef}{\begin{figure}} 
\newcommand{\ef}{\end{figure}}
\newcommand{\ie}{{\it i.e.}}
\newcommand{\eg}{{\it e.g.}}
\newcommand{\etal}{{\it et al.\ }}
\newcommand{\llabel}[1]{\label{#1}} 
\newcommand{\eq}[1]{Eq.~(\ref{#1})} 
\newcommand{\fig}[1]{Fig.~\ref{#1}} 

\title{Direct determination of the electron-electron
mean free path in diffusive mesoscopic samples using shot noise}
\author{Yehuda Naveh}
\address{Department of Physics and Astronomy, State University 
of New York, Stony Brook, NY 11794-3800}
\date{\today }
\maketitle

\begin{abstract}
Using the 'drift-diffusion-Langevin' equation, we have recently shown
that finite-frequency shot noise in diffusive mesoscopic conductors is
very sensitive 
to the ratio $\gamma \equiv L/\lee$ between the sample length $L$ and the
electron-electron mean free path $\lee$. In this work we present numerical
calculations of the noise at arbitrary value of $\gamma$. If coupled
with accurate noise measurements, the results presented here could
serve as a new and independent way of determining $\lee$ in a given
sample.
\end{abstract}

\begin{multicols}{2}

Electron-electron (e-e) scattering and the mean free path $\lee$
associated with it are among the most useful concepts in our
understanding of interacting electron systems. In contrast,
measurement schemes of $\lee$ may prove to be quite subtle. The main
reason for this is that ordinary (as opposed to umklapp)
electron-electron scattering does not change the total momentum of the
scattered electrons, and thus does not affect the conductivity. The
most widespread method to measure $\lee$ in diffusive mesoscopic (\ie,
shorter than the electron-phonon mean free path) conductors is by
fitting magneto-conductance data to weak-localization theories (see,
\eg, Ref. \cite{Altshuler 85}). This method has two fundamental
drawbacks. First, the measured effect is inherently small, \ie, it is
of the order of the quantum conductance $e^2/h$ which is much smaller
than typical conductance of the sample. Second, weak localization
measurements do not measure $\lee$ directly, but rather the dephasing
length of the electrons. At some specific circumstances (even at low
temperatures) the two lengths may not be the same\cite{Altshuler 98}.

The most direct effect e-e scattering has is on the distribution
function of electrons. Direct measurements of this distribution
function were recently performed\cite{Pothier 97} in a superconducting
tunneling spectroscopy experiment.  Such a measurement scheme, while
conceptually simple and elegant, uses intrusive lithography which may
interfere with subsequent experiments involving the same sample.  In
this paper we suggest a very different measurement scheme which may
lead to an accurate determination of $\lee$ in diffusive mesoscopic
conductors. In particular, this measurement is non-destructive to the
sample, so $\lee$ may be determined as part of the characterization of
the sample before other experiments are performed.

The measurement scheme suggested is based on shot noise measurement
in the conductor.  We have recently shown\cite{Naveh 98} that this
type of noise is very sensitive to the strength of e-e scattering in
the conductor.  For the sake of definiteness, we report in this work
results for a specific measurement geometry, namely, a geometry in
which the conductor is located close to a ground plane, with the
thickness of the conductor and the distance from the ground plane both
much smaller than the length $L$ of the conductor (The electronic
density of states in the conductor can be either three-dimensional or
two-dimensional). 
In this geometry, it was shown\cite{Naveh 98} (see Fig.~1)
that in the 
two limits 
of the 
parameter $\gamma \equiv L/l_{ee}$, two
measurable quantities acquire very different values. First, the 
spectral density of the noise $S_I(\omega)$ saturates at high
frequencies at the value $S_I(\omega)
= 0.5 \times 2eI$ ($I$ is the dc current) for $\gamma \to 0$, while it
grows as $\omega^{1/4}$ when  
$\gamma \to \infty$. Second, the temperature-dependence of
$S_I(\omega)$ 
at low lattice temperatures $T$ is linear for $\gamma \to 0$ and
quadratic for $\gamma \to \infty$. In the present work results for the
noise spectral density and its derivative with respect to temperature
are presented at intermediate values of 
$\gamma$. 

\begin{figure}[tb]
\vspace{4cm}
\centerline{\hspace{-4cm} \psfig{figure=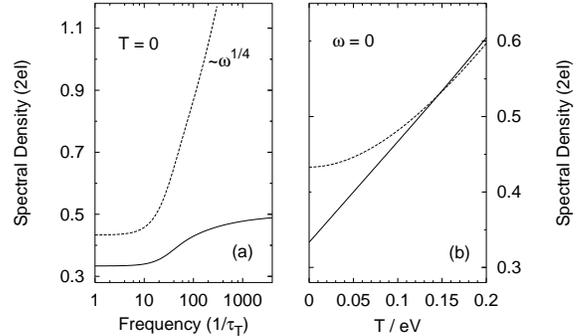,angle=-90,width=35mm}}
\narrowtext
\vspace{0.5cm}
\caption{(a) Frequency and (b) temperature dependence of the noise spectral
density for the two limiting values of the ratio $\gamma =
L/\lee$. Solid lines: $\gamma \to 0$. Dashed lines: $\gamma \to
\infty$. After Ref.~$^4$}
\label{1noee}
\end{figure}

The basis of the noise calculations presented here is the
'drift-diffusion-Langevin' theory formulated in Refs.~\cite{Naveh
98,Naveh 97}. It is based on self-consistent solution of the
Boltzmann-Langevin equation\cite{Kogan 69} (integrated over electron
momenta) together with the Poisson equation that accounts for
screening in the system (the importance of screening in affecting the
noise properties was first discussed by Landauer\cite{Landauer
95+96}). This procedure is valid at frequencies lower than $1/\tau$
and $eV/\hbar$, with $\tau$ the elastic scattering time and $V$ the
applied voltage. It enables one to calculate noise power at
frequencies comparable to the inverse Thouless time $\tau_T^{-1}$ of
electron diffusion across the sample.

The outcome of the 'drift-diffusion-Langevin' approach
may be summarized by the following simple recipe: for a conductor with
uniform cross-section of area $A$, the spectral density of current
fluctuations at any 
point $x$ (along the conductor's length) is 
\[
\label{generalnoise}S_I(x;\omega ,T)=\frac{2A}{L^2}\int_{-\frac L2}^{\frac L2%
}|K(x,x^{\prime };\omega )|^2{\cal S}(x^{\prime };T)\,dx^{\prime }.
\]
The local noise correlator ${\cal S}(x;T)$
is given by 
\[
\label{SS}{\cal S}(x;T)=2\sigma (x)\int_0^\infty dE\,f_s(E,x;T)\left[ 1-{f}%
_s(E,x;T)\right] 
\]
with $\sigma (x)$ the local conductivity and $f_s(E,x;T)$ the
momentum-symmetric part of the local (steady state) distribution function at
a total energy $E$. The response function $K(x,x^{\prime };\omega )$ equals
1 at zero frequency, but at finite frequencies it is dependent on the
specific geometry of the conductor and its electrodynamic environment,
always obeying the following sum rule: 
\[
\label{intK}\frac 1L\int_{-\frac L2}^{\frac L2}K(x,x^{\prime };\omega
)\,dx^{\prime }=1.
\]
In the geometry studied the response function
for noise current in the external electrodes assumes a  very simple
form\cite{Naveh 98}:
\[
\label{Kegrpl}K^e(x^{\prime };\omega )=\kappa \frac L2\frac{\cosh (\kappa
x^{\prime })}{{\rm sinh}(\kappa L/2)}.
\]
Here $\kappa =\sqrt{-i\omega /D^{\prime }}$, $D^{\prime }=D+\sigma A/C_0$,
$C_0$ is the (dimensionless) linear capacitance between the conductor and
the ground plane, and $D$ is the diffusion coefficient. The
non-equilibrium distribution function $%
f_s(E,x)$ should be found by solving the stationary Boltzmann equation. At
this stage it is convenient to use dimensionless quantities $\xi =x/L$, $%
\varepsilon =E/eV$, and $t=T/eV$. In the diffusion limit ($l\ll L$,
with $l$ the elastic mean free path), the
Boltzmann equation is 
\[
\label{Boltzmann}- (D / L^2) \frac{d^2f_s(\varepsilon ,\xi )}{d\xi ^2}%
=I(\varepsilon ,\xi )
\]
with $I(\varepsilon ,\xi )$ the collision integral.

The form of the collision integral for electron-electron scattering in
diffusive conductors is not agreed upon. In the most simple approach,
the electron wave functions are treated as plane waves. Then the
collision integral assumes the familiar form\cite{Altshuler
85,Gantmakher 87} 
\bea
\label{collint} \nonumber  
I(\varepsilon ,\xi ) & = & \frac 1{\tau^{ee}_V}\int d\varepsilon'
\, \int d\omega_0
\, \left[ (1-f_s) f_s' f_s^+ (1 - f_s'^{+}) \right. \\
& & \left. -f_s f_s' (1 - f_s^-) (1 - f_s'^{+} \right]
\eea
with $f_s = f_s(\varepsilon)$, $f_s' = f_s(\varepsilon')$, $f_s^\pm =
f_s(\varepsilon \pm \omega_0)$ and $f_s'^\pm = f_s(\varepsilon' \pm
\omega_0)$.  $\tau^{ee}_V \propto V^{-2}$ is the electron-electron
energy relaxation time of an electron with excess energy $eV$.  The
above form of the collision integral is strictly valid only at high
enough voltages\cite{Fukuyama 83}, $eV \gg \hbar/\tau$. At lower
voltages weak localization effects may become important. Then, the
theory predicts\cite{Altshuler 85,Fukuyama 83} $\tau^{ee}_V \propto
V^{-d/2}$ ($d$ is the dimension), and the integrand in \eq{collint}
depends explicitly on the transfered energy $\omega_0$. However, not even
a general form of the collision integral in this regime is  
known. In all works we are aware of (see, \eg, Ref.~\cite{Altshuler
85}), only the e-e relaxation rate is 
calculated. This rate may not be directly related to the collision
integral for the strongly non-equilibrium case, since it involves
only the 'scattering out' term, and this only at small deviations from
equilibrium. Indeed, the only relevant non-equilibrium experiment reported
to date\cite{Pothier 97} found results which are strongly inconsistent
with the 
relaxation 
rate of Ref.~\cite{Altshuler 85}. For these reasons 
we restrict the
present work to
the form given by \eq{collint}. However, it will be shown below that
the main qualitative 
results of the work are not dependent upon the exact form of the
collision integral $I$.

The boundary
conditions to be used with \eq{Boltzmann} are derived from the fact
that the voltage drops entirely on the sample, and therefore the
distribution function at the conductor-electrode interfaces must be
equilibrium,
\[
\label{bc}f_s(\varepsilon ,\mp 1/2)=f_0(\varepsilon \mp 1/2)\equiv \frac 1{%
1+\exp \left( \frac{\varepsilon \mp 1/2}t\right) }.
\]

As in the case of electron-phonon scattering\cite{Naveh 98a}, one can
see from equations (\ref{Boltzmann}--\ref{bc}) that 
the dependence of $f_s(E,x)$ on the physical
variables of the problem $eV$, $T$, $L$, and $\lee = \sqrt{D
\tau^{ee}_V}$ is only through the parameters $t=T/eV$ and $\gamma
=L/\lee$. From equations (\ref{generalnoise}--\ref{Kegrpl})
it is then seen that for a uniform conductor [$\sigma (x)=\sigma $] the only
additional parameter which affects the normalized noise value $\alpha
=S_I/2eI$ is $|\kappa |L=\sqrt{\omega \tau _T}$, with $\tau _T=L^2/D^{\prime
}$ the effective Thouless time. 

Equation~(\ref{Boltzmann}) was previously solved analytically in the
two limiting values of the parameter $\gamma$\cite{Nagaev 92,Nagaev
95,Kozub 96}:  The 
distribution functions in these cases are 
\begin{mathletters}
\label{limits}
\begin{eqnarray}
f_s(\varepsilon ,\xi ) &=&(1/2+\xi )f_0(\varepsilon 
+1/2) \nonumber \\
& &  +(1/2-\xi
)f_0(\varepsilon -1/2)\hspace{0.4cm} \gamma \to 0,  \label{limitsa} \\
f_s(\varepsilon ,\xi ) &=&\left\{ 1+\exp \left[ \frac{\varepsilon +\xi }{%
t_h(\xi)}\right] \right\} ^{-1}\hspace{0.4cm} \gamma \rightarrow
\infty ,  \label{limitsb}
\end{eqnarray}
\end{mathletters}
with the hot-electron temperature $t_h(\xi)=\sqrt{t^2+3(1-4\xi
^2)/4\pi ^2}$. Analytical expressions for the 
frequency and temperature dependences of the noise power in these limiting
cases were given elsewhere\cite{Naveh 98}. The main relevant results
of that 
work are summarized in Fig.~1. 
The very different
functional behavior of the noise in the two limits of $\gamma$ (which
is what motivated 
the present work) is well understood and was explained in
Ref.~\cite{Naveh 
98}.

It is imprtant to note that the limiting distributions
Eqs.~(\ref{limits}) do not depend on the form of the collision
integral $I$. This can be seen immediately by combining equations
(\ref{Boltzmann}) and (\ref{collint}). Then, it is evident that at
$\gamma \to 0$ $f_s$ is determined by equating the LHS of
\eq{Boltzmann} (which does not depend on $I$) to zero, while at
$\gamma \to \infty$ the collision integral should vanish, which means
that, for any form of $I$, $f_s$ must be given by a Fermi-Dirac
function \eq{limitsb} ($t_h$ in this case is determined solely
from conservation of total electron energy). Thus we conclude that
even in the case of $eV$ comparable to or smaller than $\hbar/\tau$,
for which $I$ is not known and therefore cannot be studied
quantitatively, the results depicted in Fig.~1 still apply, and the
noise is still very sensitive to the ratio $\gamma$.

In order to study the crossover region ({\it i.e.}, at finite values
of $ \gamma $) we solve Eq.~(\ref{Boltzmann}) numerically, utilizing
an iterative Newton-Raphson method for the discrete \eq{Boltzmann} on
a non-uniform mesh. It was found that results are accurate enough with
a mesh size of up to 200$\times$200. Results for the distribution
functions for several values of $ \gamma $ are presented in Figure 2
(the mesh plotted is considerably sparser than the one used for the
calculations). Plots (a) and (f) correspond to the limiting cases
described by equations (\ref{limitsa}) and (\ref{limitsb}),
respectively. As seen from the figure, the step-like singularity
characteristic of the distribution at $\gamma = 0$ is retained at
distances smaller than $L/\gamma^{1/2}$ from the edges of the
sample. Due to the form of the response function (\ref{Kegrpl}) the
high frequency noise is sensitive to the distribution of electrons
only at distances $\xi \sim 1/|\kappa L|=(\omega \tau _T)^{-1/2}$ from
the edges.  Therefore, since 

\begin{figure}[tb]
\vspace{0.7cm}
\centerline{\hspace{95pt} \psfig{figure=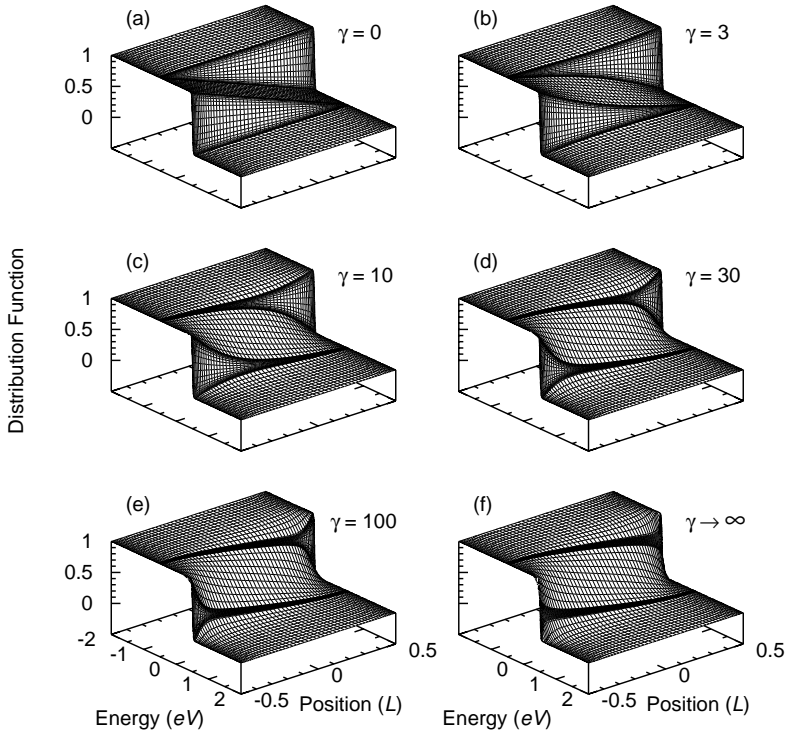,angle=-90,width=140mm}}
\narrowtext
\vspace{0.5cm}
\caption{The symmetric part of the electron distribution function
$f_s(\varepsilon 
,\xi )$ for different values of $\gamma =L/\lee$. For all curves, $%
T=0.01 \, eV$.}
\label{2distrib}
\end{figure}
\vspace{1cm}
.

\noindent 
\begin{figure}[tb]
\vspace{1.5cm}
\centerline{\hspace{-3.5cm} \psfig{figure=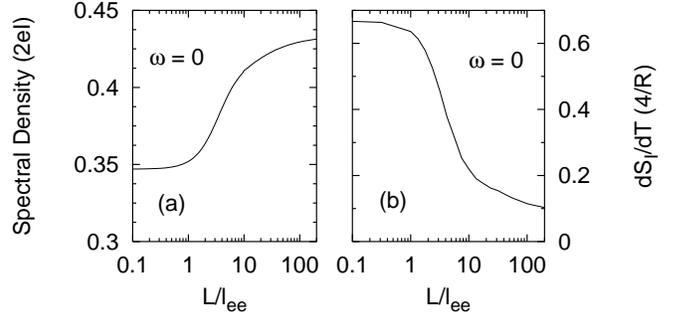,angle=-90,width=55mm}}
\narrowtext
\vspace{0cm}
\caption{The dependence of the noise spectral density (a) and its
derivative with respect to temperature (b) on $\gamma$ for
$t\equiv 
T/eV=0.01$ at zero  frequency. The units of panel (a) are normalized
to the 'full' shot noise $2eI$ while those of panel (b) are normalized
to the temperature derivative of equilibrium noise $4/R$, with $R$ the
resistance. }
\label{3zerof}
\end{figure}

\begin{figure}[tb]
\vspace{2cm}
\centerline{\hspace{-3.5cm} \psfig{figure=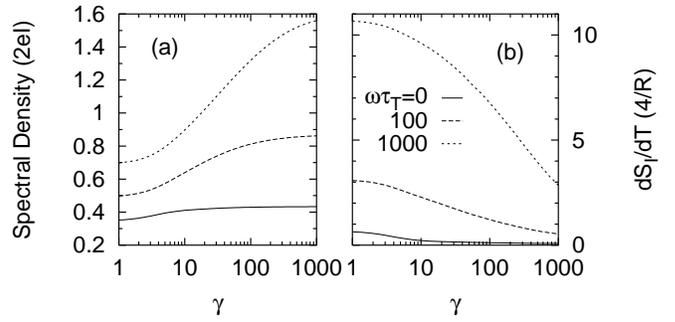,angle=-90,width=55mm}}
\narrowtext
\caption{Same as Figure 3, but for zero and finite frequencies.}
\label{4finitef}
\end{figure}

\noindent
the asymptotic behavior of the noise at
$\gamma \to \infty$ (\eg, its $\omega^{1/4}$ dependence) is due to
electron thermalization to temperature $t_h$\cite{Naveh 98}, the noise
should behave with this asymptotic form only at $ \gamma \gg \omega
\tau _T$.

Figures 3 and 4 show our main results: the normalized noise spectral
density and 
its temperature dependence as a function of $\gamma$ for various
frequencies. Fig. 3 shows the results at zero frequency, while
Fig.~4 includes also high frequencies. 
It is concluded from these figures that accurate
measurements of noise should be sensitive to the ratio $\gamma$
already at low frequencies ($\omega\tau_T \ll 1$), provided $\gamma$
is of the order of 1--10. On the other hand, the high-frequency noise
is much more sensitive to the value of $\gamma$, and can be used to
probe $\gamma$ up to a range of a few hundreds.

The possibility to use the results presented in this work as an
actual probe of $L/\lee$ relies on the advancement of accurate noise
measurements in diffusive conductors. I believe that the
present situation is very close to the necessary accuracy. Steinbach
\etal performed low frequency measurements in which 
they could unambiguously distinguish between the two limiting noise
values of Fig. 3(a). Schoelkopf \etal performed high-frequency noise
measurements with excellent accuracy. The samples in the latter
experiments were made of well-conducting gold, so the crossover
frequency $1 / \tau_T$ was very high. In samples with poorer
conductivity the 'high-frequency' regime may be reached at $100/2 \pi
\tau_T \sim 10$ GHZ, so the curves of Fig.~4 can be
experimentally verified at microwave frequencies. In addition to
absolute measurements of noise, the temperature dependence shown in
Figures 3(b), 4(b) should be relatively simple to detect.

The results presented here assume that the sample is strictly in the
mesoscopic regime, so that no thermalization by phonons
occurs. Practically, at
$\gamma > 100$ phonon relaxation may start to be appreciable even at
relatively low temperatures. Theoretically, this means that the
electron-phonon collision term should be added to the collision
integral (\ref{collint}). However, the subsequent suppression of the
noise compared to the results presented here may not be substantial
because significant reduction of the noise, particularly at high
frequencies, happens only at a very large ratio of L and the
electron-phonon relaxation length\cite{Naveh 98a}.

To summarize, we have performed numerical calculations of the
non-equilibrium noise in mesoscopic samples with strong elastic
scattering and with an arbitrary strength of electron-electron
scattering. We have shown that the spectral density of the high
frequency noise and its functional dependence on temperature are very
sensitive to the ratio $\gamma \equiv L/\lee$. The results presented
here, if coupled with accurate noise measurements, can thus serve as a
new and independent way to identify this ratio in a given sample.

The author is indebted to D.V. Averin, K. K. Likharev and D. Menashe
for many fruitful discussions.
The work was supported in part by DOE's Grant \#DE-FG02-95ER14575.

\end{multicols} 
\end{document}